\documentclass[aps,prl,twocolumn,nofootinbib,superscriptaddress]{revtex4-2}

\pdfoutput=1

\usepackage[utf8]{inputenc} 
\usepackage{amsmath}
\usepackage{amssymb}
\usepackage{xcolor}
\usepackage{graphicx}
\usepackage{dcolumn}
\usepackage{bm}
\usepackage{microtype}
\usepackage{threeparttable}
\usepackage{mathtools}
\usepackage[colorlinks=true]{hyperref}
\usepackage{physics}
\usepackage{comment}

\newcommand{\be}{\begin{eqnarray}}
\newcommand{\ee}{\end{eqnarray}}

\begin{document}

\title{Quasinormal modes of nonthermal fixed points}

\author{Matisse De Lescluze}
\email{matisse.delescluze@ugent.be}
\affiliation{Department of Physics and Astronomy, Ghent University, 9000 Ghent, Belgium}

\author{Michal P. Heller}
\email{michal.p.heller@ugent.be}
\affiliation{Department of Physics and Astronomy, Ghent University, 9000 Ghent, Belgium}

\begin{abstract}

\noindent Quasinormal modes play a prominent role in relaxation of diverse physical systems to equilibria, ranging from astrophysical black holes to tiny droplets of quark-gluon plasma at RHIC and LHC accelerators. We propose that a novel kind of quasinormal modes govern the direct approach to self-similar time evolution of nonthermal fixed points, whose relevance ranges from high energy physics to cold atom gases. We utilize black hole perturbation theory techniques to compute the spectrum of these far from equilibrium quasinormal modes for a kinetic theory with a Focker-Planck collision kernel in isotropic and homogeneous states. Our conclusion is that quasinormal modes of nonthermal fixed points give rise to a tower of progressively more decaying power-law  contributions. A byproduct of our analysis is a precise determination and improved understanding of the distribution function characterizing nonthermal fixed points.
\end{abstract}

\maketitle

\noindent \textbf{\emph{Introduction.--}} Understanding thermalization in quantum-many body systems and quantum field theories (QFTs) is an important topic of contemporary research both on the theory and experiment frontiers~\cite{Berges:2020fwq,Ueda:2020ehs}. When a QFT has many microscopic constitutents that are strongly interacting then thermalization is described in a dual gravitational description~\cite{Maldacena:1997re,Gubser:1998bc,Witten:1998qj} by black hole formation and its subsequent relaxation with time~$t$~\cite{Chesler:2008hg}. Similarly to astrophysical black holes, this relaxation is captured by an infinite tower of complex eigenfrequencies $\omega$ - quasinormal modes (QNMs)~\cite{Horowitz:1999jd,Kovtun:2005ev,Berti:2009kk}. Their imaginary part governs the decay of observables such as the expectation value of the energy-momentum tensor to their thermal values and their real part is responsible for generically present associated oscillations: $e^{-i \omega t}$. As time progresses, the least damped QNM starts dominating physical observables.

In this letter we recognize that QNMs should play another, so far unexplored role in thermalization dynamics. In the course of the past 20 years it has been predicted theoretically~\cite{Micha:2002ey,Berges:2008wm,Berges:2013eia,Kurkela:2015qoa} and then realized experimentally~\cite{Prufer:2018hto,Erne:2018gmz,Glidden:2020qmu,
Navon:2016em,johnstone2019evolution,Helmrich:2020sgn,Garcia-Orozco:2021hkx,Martirosyan:2023mml,Huh:2023xso,Lannig:2023fzf,Gazo:2023exc,Martirosyan:2024rxm} that overoccupied states are attracted to a transient albeit long-lived nonthermal fixed point (NTFP) regime characterized by a temporal self-similarity, see also~\cite{Berges:2015kfa,Schmied:2018mte} for reviews. In the simplest isotropic case, the weakly-coupled dynamics is characterized by a distribution function of particles $f(t,p = |\vec{p}|)$ depending on time $t$ and the magnitude of spatial momentum $p$. The NTFP regime occurs in this setup when the temporal and momentum dependence effectively factorize
\begin{equation}
\label{eq.NTFPdef}
f(t, p) = A(t) f_{s}(B(t) p).
\end{equation}
As a result, in the NTFP regime the form of a distribution function at a given time allows to predict its form to the future by a mere rescaling of the distribution function and momentum. The Boltzmann equation
\begin{equation}
\label{eq.Boltzmann}
    \partial_t f(t,p)=C[f](t,p),
\end{equation}
dictates then that
\begin{equation}
\label{eq.ABdef}
A(t) = B(t)^{\sigma} \quad \mathrm{and} \quad B(t) = \left(\frac{t - t_{*}}{t_{ref}}\right)^{\beta}.
\end{equation}
The number~$\sigma$ and the exponent~$\beta$ characterize the underlying dynamics, $t_{ref}$ is a constant choosing normalization of $f_{s}$ and $t_{*}$ is an initial condition dependent offset picking the origin of time with 0 playing a special role for self-similarities such as~\eqref{eq.ABdef}~\cite{Heller:2023mah,Gazo:2023exc}.

Our key insight on top of the aforementioned observed attractive nature of NTFPs is that $A(t)^{-1} f(t,p)$ becomes effectively time-independent when viewed as a function of $\bar{p} \equiv B(t) p$ and renders this configuration effectively static in a close similarly to the static nature of the thermal state or a final black hole in a merger or collapse. This allows for setting up eigenvalue problem for perturbations
\begin{equation}
\label{eq: perturbation to prescaling bis}
A(t)^{-1}f(t,\bar{p}/B(t)) \approx f_{s}(\bar{p}) + \delta f(t,\bar{p}).
\end{equation}
around a NTFP regime. The QNM ansatz turns out to be in this case
\begin{equation}
\label{eq: time dependence delta f}
\delta f(t,\bar{p}) = B(t)^{i \, \Omega} \delta f_{\Omega}(\bar{p}).
\end{equation}
The spectrum of eigenfrequencies $\Omega$ and eigenfunctions $\delta f_{\Omega}$, the QNMs of a given NTFP, is determined by linearization of the Boltzmann equation~\eqref{eq.Boltzmann} around the scaling solution~\eqref{eq.NTFPdef}.

Our general expectation is therefore that a NTFP is going to be approached at sufficiently late times in a power law manner with a possible real part of $\Omega$ triggering slow oscillations logarithmic in time. At intermediate times with deviations from the scaling form~\eqref{eq.NTFPdef} small it is going to be a sum of QNMs that describe the signal. We expect that the lessons drawn in the context of black hole QNMs, including various recent developments~\cite{Withers:2018srf,Grozdanov:2019kge,Arean:2023ejh,Cownden:2023dam,Carballo:2024kbk}, will be generally transferable to the novel setting of NTFPs.

In order to corroborate these general expectations, we explicitely compute the spectrum of QNMs in NTFPs arising for the collision kernel in the Boltzmann equation of the Fokker-Planck form, see Eqs.~\eqref{eq: CFP} and~\eqref{eq: integrals FP}, relevant for, e.g., small angle elastic scattering of gluons at very weak coupling.

\vspace{5 pt}

\noindent \textbf{\emph{QNM eigenproblem around NTFPs.--}} The emergence of exact NTFPs~\eqref{eq.NTFPdef} is a consequence of the collision kernel~$C$ in Eq.~\eqref{eq.Boltzmann} being a homogeneous functional of particle momenta~\cite{Micha:2004bv,Heller:2023mah}. Using~\eqref{eq.NTFPdef} the collision kernel must then behave as
\begin{equation}
\begin{split}
    C[f](t,p)&=A^{\mu_\alpha}(t)B(t)^{\mu_\beta}\Tilde{C}[f_s](\bar{p})\\&=B(t)^{\sigma \mu_\alpha+\mu_\beta}\Tilde{C}[f_s](\bar{p}).
    \end{split}
    \label{eq: homogeneous scaling collision kernel}
\end{equation}
We consider now a small perturbations $\delta f \ll f_s$, as in Eq.~\eqref{eq: perturbation to prescaling bis}. The parameterization of $\delta f$ was chosen to make it easy to capture time dependence in the static frame of~$f_s$. The way we parameterized the dependence  of $\delta f$ on $A(t)$ and $B(t)$ in Eq.~\eqref{eq: perturbation to prescaling bis} together with Eq.~\eqref{eq: homogeneous scaling collision kernel} implies
\begin{equation}
\begin{split}
    C[f](t,p)&=B(t)^{\sigma \mu_\alpha+\mu_\beta}\Tilde{C}[f_s](\bar{p})\\&+B(t)^{\sigma \mu_\alpha+\mu_\beta}\Tilde{C}[fs,\delta f](t,\bar{p}).
\end{split}
    \label{eq: homogeneous scaling collision kernel with perturbation bis}
\end{equation}
\noindent The other defining feature of NTFPs, i.e. conservation of energy or particle number, is ensured by relating $A(t)$ and $B(t)$ by a positive constant~$\sigma$~\cite{Heller:2023mah}, see Eq.~\eqref{eq.ABdef}. Conservation laws also constrain the behavior of perturbations around a NTFP, as will be explained in a later section. We apply linear response theory by plugging the form of $f(t,p)$ in Eq.~\eqref{eq: perturbation to prescaling bis} into the Boltzmann equation~\eqref{eq.Boltzmann} and considering only terms up to first order in $\delta f$. By making use of Eqs.~\eqref{eq: homogeneous scaling collision kernel with perturbation bis} and ~\eqref{eq.ABdef}, we obtain
\begin{equation}
\frac{B(t)^{\sigma \mu_\alpha+\mu_\beta}}{B(t)^{\sigma-1}\partial_tB(t)}=\frac{1}{D_1}=\frac{[\sigma+\bar{p}\cdot\partial_{\bar{p}}]f_s(\bar{p})}{\tilde{C}[f_s](\bar{p})},
    \label{eq: seperation of variables bis}
\end{equation}
\begin{equation}
    \begin{split}
        &B(t) \partial_{B(t)} \delta f\left(t,\bar{p}\right)|_{\bar{p}=const}\\
        &=\frac{1}{D_1}\Tilde{C}[f_s,\delta f](t,\bar{p})-\sigma \delta f\left(t,\bar{p}\right)- \bar{p}\partial_{\bar{p}}\delta f\left(t,\bar{p}\right),
    \end{split}
    \label{eq: equation for delta f bis}
\end{equation} 
with  $D_1$ a separation of variables constant. Eq.~\eqref{eq: seperation of variables bis} was studied in \cite{Heller:2023mah} with the left equality determining the form of~$B(t)$. The solution is given by Eq.~\eqref{eq.ABdef} with~$D_1=\frac{\beta}{t_{ref}}$ and $1/\beta=(1-\mu_\alpha)\sigma-\mu_\beta$. We used~Eq.~\eqref{eq: seperation of variables bis} to write the equation for $\delta f$ in the form~\eqref{eq: equation for delta f bis}. The latter is solved by the Ansatz~\eqref{eq: time dependence delta f}. The sign convention for the imaginary unit was chosen so that later in the paper the decaying modes will be characterized by $\Im(\Omega)<0$ as is the usual convention. Because $\Omega$ is constant we can pull the $B(t)^{i \Omega}$ factor through the operators on the right-hand side and eliminate it leading to
\begin{equation}
\begin{split}
    &i \Omega \delta f_\Omega(\bar{p})\\&=\frac{1}{D_1}\Tilde{C}[f_s,\delta f_\Omega](\bar{p})-\sigma \delta f_\Omega\left(\bar{p}\right)-\bar{p}\partial_{\bar{p}}\delta f_\Omega\left(\bar{p}\right).
\end{split}
\label{eq: eigenvalue equation}
\end{equation}
Eqs.~\eqref{eq: time dependence delta f} and \eqref{eq: eigenvalue equation} are the main results of this paper. They show that perturbations on top of a NTFP behave in the same way in time as the self-similar distribution function, but with different exponents $(\sigma+i \Omega) \beta$, see also Eq.~\eqref{eq: perturbation to prescaling bis}. The perturbations through their imaginary part induce a power law in time approach to~\eqref{eq.NTFPdef} when $\Re(i \Omega) \beta < 0$, which is a self-consistency condition for the attractive nature of NTFPs observed in ab initio simulations and experiments. Furthermore, real parts of $\Omega$'s, if nonzero, would induce oscillatory behaviour in the logarithm of~$B$, i.e. in $\log(t-t_{\star})$. This is reminiscent of the near equilibrium kinetic theory setup of~\cite{Heller:2018qvh}.

The last key aspect that we want to highlight is that the equation for perturbations becomes naturally independent of the value of the coupling constant at the leading order in the coupling, as well as on $D_{1}$ and $t_{ref}$ sitting therein. One can see it by comparing the structures of the second equality in Eq.~\eqref{eq: seperation of variables bis} with the combination $\frac{1}{D_1}\Tilde{C}[f_s,\delta f_\Omega](\bar{p})$ in Eq.~\eqref{eq: eigenvalue equation}. In practical terms, this implies that at small couplings the QNM frequencies naturally come out independent of the coupling.

\vspace{5 pt}

\noindent \textbf{\emph{Two universally present QNMs.--}} A corroboration of the discussion above is the existence of two special QNMs~\eqref{eq: time dependence delta f} around \emph{any} NTFP. Looking back at Eqs.~\eqref{eq.NTFPdef} and~\eqref{eq.ABdef} there are two parameters which we can perturb and investigate the response of the system, namely $t_\ast$ and~$t_{ref}$. 

Let us first consider $t_\ast\rightarrow t_\ast+\delta t_\ast$ in Eqs.~\eqref{eq.NTFPdef} and~\eqref{eq.ABdef} and expand to first order in $\delta t_\ast$
\begin{equation}
\begin{split}    f(t,p)&=A(t)f_s(\bar{p})\\&-\frac{\beta}{t_{ref}}A(t)B(t)^{-1/\beta}\left(\sigma f_s(\bar{p})+\bar{p}\partial_{\bar{p}}f_s(\bar{p})\right).
\end{split}
    \label{eq: ts mode}
\end{equation}
This has the same structure as predicted in \eqref{eq: perturbation to prescaling bis} and \eqref{eq: time dependence delta f} with $\Omega=\frac{i}{\beta}$ and $\delta f_\Omega(\bar{p})\propto\sigma f_s(\bar{p})+\bar{p}\partial_{\bar{p}}f_s(\bar{p})$. One can think of this perturbation as arising at very late times when $t_{*}$ is not included in $A$ and $B$, as was the case before~\cite{Heller:2023mah,Gazo:2023exc}.

The perturbation of $t_{ref}$ is more subtle. We can repeat the steps above when considering equations \eqref{eq: perturbation to prescaling bis} and \eqref{eq: time dependence delta f}. This would lead to $\Omega=0$ and the same momentum dependence as for the $t_\ast$ perturbation. This, of course, makes no sense since there would be two different eigenvalues for the same eigenfunction. One has to be more careful and realise that $f_s$ itself also contains a non-trivial dependence on $t_{ref}$, see Eq.~\eqref{eq: seperation of variables bis} with $D_1=\frac{\beta}{t_{ref}}$. The correct form of $f(t,p)$ for $t_{ref}\rightarrow t_{ref}+\delta t_{ref}$ is
\begin{equation}
\begin{split}
    f(t,p)&=A(t)f_s(\bar{p})\\&-\frac{\beta}{t_{ref}}A(t)\left(\sigma f_s(\bar{p})+\bar{p}\partial_{\bar{p}}f_s(\bar{p})\right)\\&+A(t)\partial_{t_{ref}}f_s(\bar{p})|_{\bar{p}=const}
\end{split}
    \label{eq: tref mode}
\end{equation}
\noindent We found a zero-mode $\Omega=0$ with momentum dependence $\delta f_\Omega(\bar{p})\propto\sigma f_s(\bar{p})+\bar{p}\partial_{\bar{p}}f_s(\bar{p})-\frac{t_{ref}}{\beta}\partial_{t_{ref}}f_s(\bar{p})|_{\bar{p}=const}$. This perturbation is the analog of a variation with respect to thermodynamic variables like temperature or chemical potential in equilibrium~\cite{Denicol:2022bsq}. The fact that shifting $t_{ref}$ leads to a zero-mode can be easily understood. The reference time is a dummy constant and fixes the normalisation of $f_s$. Using a different $t_{ref}$ leads to a different scaling function in Eq.~\eqref{eq.NTFPdef}. The NTFP form is retained and in the static frame of the scaling function nothing decays away.

\vspace{5 pt}

\noindent \textbf{QNMs and conservation laws.--} There are a several more important statements to make about NTFP's QNMs in full generality. First we discuss the importance of conservation laws. The dynamics of self-similar cascades for systems reaching a NTFP, is dictated by conserved quantities, e.g. particle number density $n=\int d^dp f/(2\pi)^d$ and energy density $\epsilon=\int d^dp \omega_{\vec{p}}
f/(2\pi)^d$. With $d$ the number of spatial dimensions and $\omega_{\vec{p}}$ is the
dispersion relation of particles. We focus on $\omega_{\vec{p}}\propto|\vec{p}|^z=p^z$, as in \cite{Heller:2018qvh}. The relation between $A(t)$ and $B(t)$ in Eq.~\eqref{eq.ABdef} is the consequence of having one conserved quantity. Conservation of $n$ leads to $\sigma=d$, while conservation of $\epsilon$ corresponds to $\sigma=d+z$. This only imposes the relevant conserved quantity to be constant, when caclulated from $f_s$ in the static frame $\bar{p}$. We define the conserved quantity to be completely contained by $f_s$. Considering the example of energy density, this implies

\begin{equation}
    \epsilon=\int d^dp p^z f(t,p)=\int d^d\bar{p}\bar{p}^z f_s(\bar{p}).
\end{equation}

\noindent One can always ensure this by choosing $t_{ref}$ in such a way that the above equation holds. It is now clear that~$\epsilon$ can only be a constant when the energy density of the perturbations defined in Eq.~\eqref{eq: perturbation to prescaling bis} in the static frame of $f_s$ is time-independent

\begin{equation}
\begin{split}
         \delta\epsilon(t)&=\int d^d\bar{p}\bar{p}^z \delta f(t,\bar{p})\\&=B(t)^{i\Omega}\int d^d\bar{p}\bar{p}^z \delta f_\Omega(\bar{p})=const.
\end{split}
\end{equation}

\noindent There are two ways for the equation above to be constant. Either $\Omega=0$ or $\delta\epsilon_\Omega\equiv\int d^d\bar{p}\bar{p}^z \delta f_\Omega(\bar{p})=0$. Both of these conditions are already clear from our predicted modes. As explained, the zero-mode corresponds to a shift in $t_{ref}$. But the choice of $t_{ref}$ ensures that $f_s$ contains the same energy density as $f$. Changing $t_{ref}$ thus changes the overall energy density, which is a different but allowed solution of the equations of motion near a NTFP. Because in homogeneous and isotropic case there is only one conserved quantity dictating the dynamics, we only expect there to be one zero-mode. The $t_\ast$ mode, for $\sigma=d+z$, makes the other condition $\delta\epsilon_\Omega=0$ apparent. 

\begin{equation}
\begin{split}
    \delta\epsilon_\Omega&\propto\int d\bar{p} \bar{p}^{d+z}\left\{(d+z)) f_s(\bar{p})+\bar{p}\partial_{\bar{p}}f_s(\bar{p})\right\}\\
    &=\int d\bar{p}\partial_{\bar{p}}\left(\bar{p}^{d+z}f_s(\bar{p})\right),
\end{split}
\label{eq: energy conservation ts mode}
\end{equation}

\noindent which is zero if $\bar{p}^{d+z}f_s(\bar{p})$ vanishes at the boundaries of the integral. We discuss the effect of boundaries in more depth in the section where the scaling function is discussed. The discussion is completely analogous when considering conservation of particle number density.

We want to highlight that these ideas are completely general and do not assume specific collision kernels. Importantly, we did not make any additional assumptions to the ones that are needed to realise prescaling. Next, we proceed to test these general lessons in the explicit example of the collision kernel of the Fokker-Planck type.

\vspace{5 pt}

\noindent \textbf{\emph{Setup.--}} The Fokker-Planck kinetic theory adopted from QCD applications is given by
\begin{equation}
\begin{split}
    C_{FP}[f](t,p)&=K\left\{\frac{I_a}{p^2}\partial_p\left(p^2\partial_p f\right)\right.\\&+\left.\frac{I_b}{p^2}\partial_p\left(p^2 f \left(f+1\right)\right)\right\}
\end{split}
\label{eq: CFP}
\end{equation}
\begin{equation}
\begin{split}
    &K=\frac{\lambda^2}{4\pi}\mathcal{L}=\frac{\lambda^2}{4\pi}\\
    &I_a=\int\frac{d^3p}{(2\pi)^3}f(f+1)\\
    &I_b=2 \int\frac{d^3p}{(2\pi)^3}\frac{f}{p},\\
\end{split}
\label{eq: integrals FP}
\end{equation}
see e.g.~\cite{Berges:2020fwq}. In $K$, $\lambda$ is the ('t Hooft) coupling constant and the Coulomb logarithm $\mathcal{L}$ is normally given by $\log(\left<p\right>/m_D)$, but we chose to set it equal to $1$. In other words, we assume the argument of the logarithm to be independent of time. Doing this makes the FP kernel scale as Eq.~\eqref{eq: homogeneous scaling collision kernel}. If the time dependence in~$K$ would be taken into account, it would induce small, slowly decaying logarithmic corrections to~\eqref{eq.ABdef}~\cite{Brewer:2022vkq,Heller:2023mah}. This effect can be included in perturbation theory around the solutions we provide, which we leave for future investigations.

\noindent When assuming overoccupation and in the static frame of $f_s$ the FP kernel reduces to:
\begin{equation}
\begin{split}
    \Tilde{C}_{FP}[f_s](\bar{p})&=K\left\{\frac{\Tilde{I}_a}{\bar{p}^2}\partial_{\bar{p}}\left(\bar{p}^2\partial_{\bar{p}} f_s\right)\right.+\left.\frac{\Tilde{I}_b}{\bar{p}^2}\partial_{\bar{p}}\left(\bar{p}^2 f_s^2 \right)\right\}
\end{split}
\label{eq: CFP overoccupied}
\end{equation}
\noindent with 
\begin{equation}
\begin{split}
    &\Tilde{I}_a=\int\frac{d\bar{p} \bar{p}^2}{2\pi^2}f_s^2\\
    &\Tilde{I}_b=\int\frac{d\bar{p} \bar{p}}{\pi^2}f_s.\\
\end{split}
\label{eq: integrals FP overoccupied}
\end{equation}
Conservation of energy density $\epsilon$ leads to $\sigma=4$ and the direct energy cascade NTFP $\left(\alpha,\beta\right)=\left(-4/7,-1/7\right)$.

Our goal is to solve Eq.~\eqref{eq: eigenvalue equation}, to this end we first need an accurate determination of the scaling function~$f_s$. In order to obtain both, we will descretize the momentum $\bar{p}$ using using a Chebyshev-Gauss-Lobatto pseudospectral method~\cite{Boyd:Chebyshev}. It is based on an expansion in basis functions, namely the Chebyshev polynomials $T_N(x)=\cos\left(N\cos^{-1}(x)\right)$ defined on the interval $\left[-1,1\right]$. This interval is discretized as the nodes $x_k=-\cos(\frac{k\pi}{N})$ of $T_N$, for $k=0,...,N$. Now, a general function can be represented as a column vector with each argument being its value at the corresponding gridpoint. By expressing the derivative in the basis of Chebyshev polynomials, a discretized derivative and integral operator can be defined. Their expressions are known and can be found in e.g. \cite{Boyd:Chebyshev}. These numerical operators will be represented by matrices. Rescaling the grid points and operators with the correct factors allows this method to be used for an arbitrary momentum grid $\bar{p}=[\bar{p}_{IR},\bar{p}_{UV}]$. Because pseudospectral methods are based on expansions in basis functions defined on the full functional momentum grid, they take into account the full structure of functions when taking derivatives. This lead to very accurate results for derivatives and integrals for less gridpoints compared to finite difference method. The matrices representing operators will however be full matrices instead of the usual sparse matrices for finite differences. See, e.g.,~\cite{Janik:2015waa} for the pseudospectral approach applied to determine QNMs of holographic black holes.

\vspace{5 pt}

\noindent \textbf{\emph{Accurate determination of the scaling function.--}} We outline now a new method to determine $f_s$ by directly solving the right equality in Eq.~\eqref{eq: seperation of variables bis} with  technical aspects in the Appendix~3. From Eqs.~\eqref{eq: seperation of variables bis}, \eqref{eq: CFP overoccupied} and \eqref{eq: integrals FP overoccupied} it is clear that $f_s$ will depend self-consistently on $\tilde{I}_a$, $\tilde{I}_b$ and $K$. By rescaling momentum and $f_s$
\begin{equation}
         \bar{p} \rightarrow \sqrt{\tilde{I}_a D_2}\bar{q} \quad \mathrm{and}\quad f_s(\bar{p})\rightarrow \frac{\sqrt{\tilde{I}_a}}{\tilde{I}_b \sqrt{D_2}} g_s(\bar{q}),
\label{eq: subsitutions}
\end{equation}
where $D_2=K \left|D_1\right|$, we get rid of $\tilde{I}_a$, $\tilde{I}_b$ and $K$ in the equation. This also makes it clear how they appear in~$f_s$. The equation that determines $g_s$ and how it is solved are explained in the Appendix~3. To find a solution that corresponds to an allowed scaling function, we constrain~$g_s$ in the following way. Making the same substitutions as in \eqref{eq: subsitutions} in the expressions for $\tilde{I}_a$ and $\tilde{I}_b$ results in 
\begin{equation}
\begin{split}
    &\tilde{I}_a=\frac{\tilde{I}_a^{5/2} \sqrt{D_2}}{\tilde{I}_b^2}\int\frac{d\bar{q} \bar{q}^2}{2\pi^2}g_s^2,\\
    &\tilde{I}_b=\frac{\tilde{I}_a^{3/2} \sqrt{D_2}}{\tilde{I}_b}\int\frac{d\bar{q} \bar{q}}{\pi^2}g_s.\\
\end{split}   
\label{eq: Ia and Ib from gs}
\end{equation}

\noindent Dividing both equations, eleminates $\tilde{I}_a$, $\tilde{I}_b$ and $D_2$. The remaining equation is $\tilde{I}_a[g_s]\equiv\int\frac{d\bar{q} \bar{q}^2}{2\pi^2}g_s^2=\int\frac{d\bar{q} \bar{q}}{\pi^2}g_s\equiv\tilde{I}_b[g_s]$.
It is clear that we can construct a self-consistent scaling solution from $g_s$ if $\tilde{I}_a[g_s]=\tilde{I}_b[g_s]$. Then $\tilde{I}_a$ and $\tilde{I}_b$ are related as in the equations above. This means we can choose either the value for $\tilde{I}_a$ or $\tilde{I}_b$ arbitrarily and the other follows from Eq.~\eqref{eq: Ia and Ib from gs}. To keep life easy we choose $\tilde{I}_a=1/D_2$, which means $\bar{q}\equiv \bar{p}$ so we do not have to rescale our numerical grids and we will simply keep using~$\bar{p}$. Utilizing $\tilde{I}_a[g_s]$ to calculate $\tilde{I}_b$, as in the equation above, means $\tilde{I}_a=1/D_2$ is exact. Finally, we get a solution for the scaling function $f_s$, after rescaling $g_s$ by the amplitude $1/\sqrt{D_2\tilde{I}_a[g_s]}$, from Eqs.~\eqref{eq: subsitutions} and~\eqref{eq: Ia and Ib from gs} with $\tilde{I}_a=1/D_2$.
\begin{figure}[h!]
    \centering
    \includegraphics[width=0.9\linewidth]{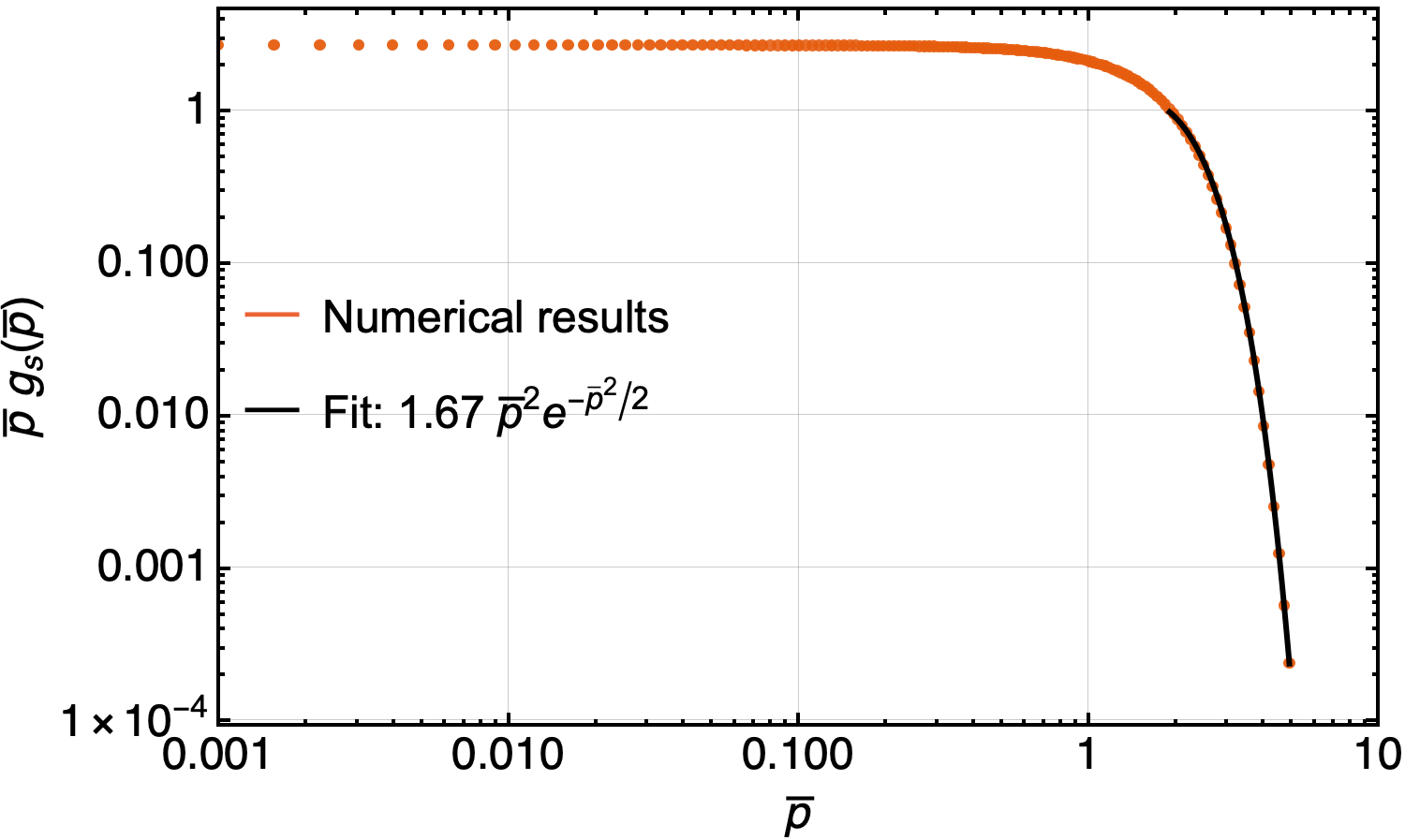}
    \caption{The plot of $\bar{p} g_s(\bar{p})$ calculated without cut-offs on the compact grid $u=\frac{\bar{p}}{1+\bar{p}}$. The plot of $\bar{p}\,f_s$ will look the same with the~$y$ axis multiplied by $\bar{p}^{-1}/\sqrt{D_2\tilde{I}_a[g_s]}$. The red curve is a fit to the UV tail, consistent with the Gaussian behavior mentioned in the main text. The last 55 data points were dropped but the Gaussian tail extends all the way to the last gridpoint.\label{fig: scaling function}}
\end{figure}
\noindent To probe arbitrarily large momenta, we introduce a compactified variable $u=\frac{\bar{p}}{1+\bar{p}}$. This way $\bar{p}=0$ coincides with $u=0$ and $\bar{p}=\infty$ coincides with $u=1$. For a system without cut-offs, $u$ is numerically represented by a compact grid $[0,1]$ as explained in the Appendix~3. The solution $g_s$ that satisfies $\tilde{I}_a[g_s]=\tilde{I}_b[g_s]$ most accurately on this compact grid goes as $\propto 1/\bar{p}$ for $\bar{p}\ll1$ and $\propto \bar{p} e^{-\bar{p}^2/2}$ for $\bar{p}\gg 1$. The numerical solution obtained iteratively, see the Appendix~3, is shown in Fig.~\ref{fig: scaling function}. The fact that the UV tail is Gaussian, $\propto \bar{p} \, e^{-\bar{p}^2/2}$, is not as straightforward as one might think. For large $\bar{p}$ the leading behavior from the equation for $g_s$ is actually $\propto1/\bar{p}^4$. The constraint that makes the integrals for $\tilde{I}_a$ and $\tilde{I}_b$ conistent with their appearance in the overoccupied FP kernel, \eqref{eq: CFP overoccupied}, selects the solution that has a Gaussian tail. This is shown in all details in Appendix~3. Interestingly, a power law tail $\propto1/\bar{p}^4$ would lead to an infinite energy density for a momentum grid without cut-offs and would thus be physically unrealisable. Ref.~\cite{AbraaoYork:2014hbk} found almost the same behavior for a scaling function resulting from ab initio simulations using QCD effective kinetic theory~\cite{Arnold:2002zm}. They see $1/\bar{p}$ in the IR up to the momentum scale where the kinetic description is valid and fit $\frac{1}{\bar{p}}(a e^{-b\bar{p}}+ce^{-d\bar{p}^2})$ in the UV. Their values for the fit parameters actually make the Gaussian part the only significantly contributing term for the UV momenta they consider. The FP kernel is an approximation of only one of the terms in the kinetic theory used in \cite{AbraaoYork:2014hbk}. The fact that we  also find a Gaussian tail using a very different method from~\cite{AbraaoYork:2014hbk} provides a nontrivial corroboration of our solution.

\begin{figure}[h]
    \centering
    \includegraphics[width=0.95\linewidth]{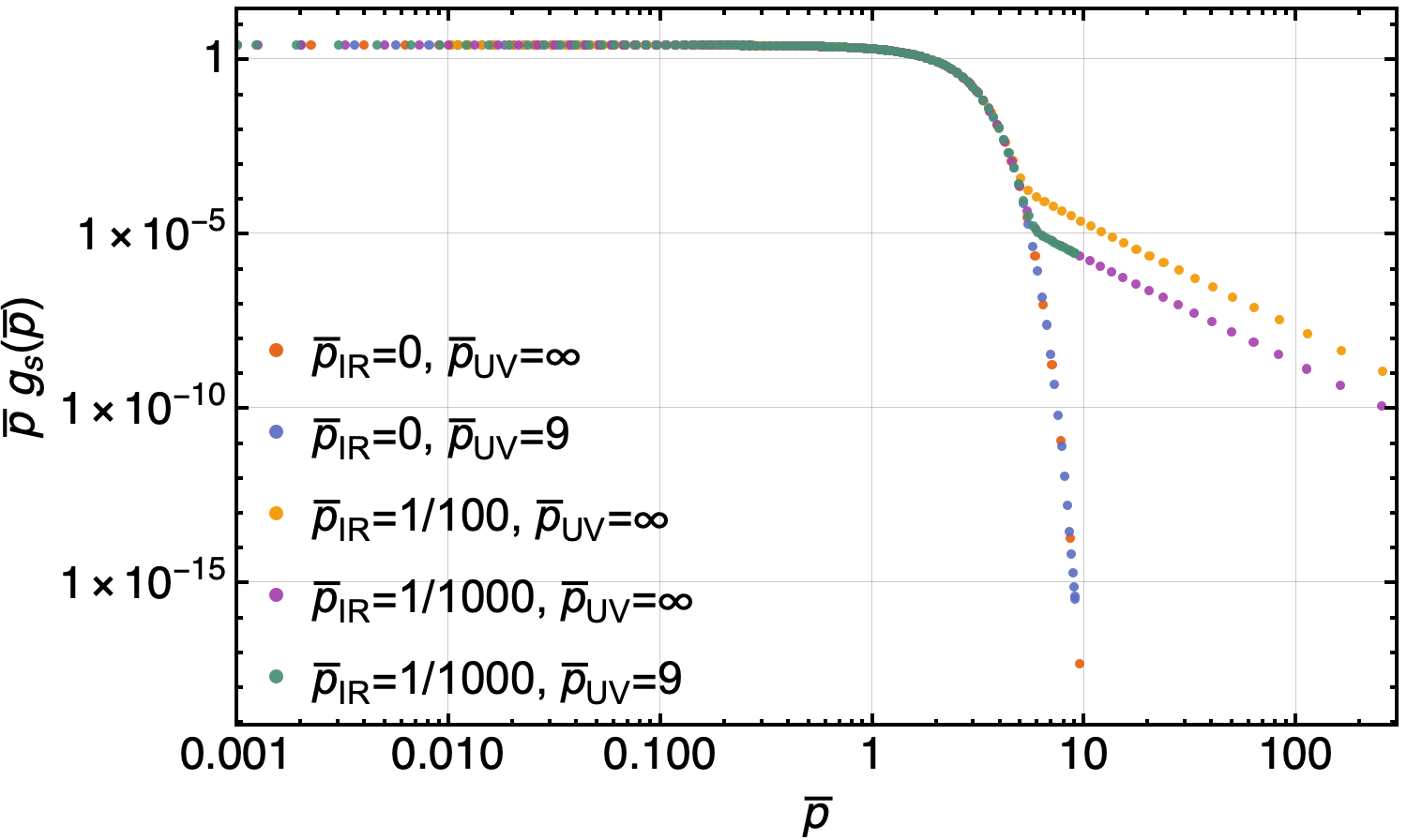}
    \caption{This plot depicts the dependence of the scaling function on UV and IR cut-offs. One sees that only UV cut-off presence does not alter the behaviour significantly in comparison to Fig.~\ref{fig: scaling function}. However, IR cut-off gets rid of the Gaussian tail and replaces it by a powerlaw $\propto1/\bar{p}^4$ visible as datapoints alinging along straight lines. The coefficient of the powerlaw is IR cut-off dependent.}
    \label{fig: scaling function cut-offs}
\end{figure}

Let us now discuss the effect of cut-offs. For cut-offs on either side of the momentum grid, the IR behavior of $g_s$ remains the same, i.e. $\propto 1/\bar{p}$, see Fig.~\ref{fig: scaling function cut-offs}. Introducing a UV cut-off leads to no significant changes in the limiting behavior of $g_s$. This is a consequence of the fact that the distribution function is already so small for large enough values of $\bar{p}$, because of the Gaussian tail, that dropping a part of the UV tail does not influence the integrals for $\tilde{I}_a[g_s]$ and $\tilde{I}_b[g_s]$ significantly. For an IR cut-off, the story is different and the $\bar{p} e^{-\bar{p}^2/2}$ seems to go over into a power-law tail $\propto 1/\bar{p}^4$. The $\bar{p}$ value where the Gaussian tail goes over into a power-law grows when the IR cut-off is decreased, see Fig.~\ref{fig: scaling function cut-offs}. A UV tail $\propto1/\bar{p}^4$ leads to divergent energy densities and it is unclear how physical processes would result in such a distribution in the UV. This clearly signals an unphysicality and to alleviate this we also include a UV cut-off. This is a very natural procedure in the formalism to describe cascades in systems at a NTFP. The assumption of overoccupation can not hold for all times over an infinte momentum range for a finite system. For large momenta the occupation will drop and the tail will eventually thermalise. The idea is then that the assumption holds between two cut-offs between which there is an approximate conserved quantity. This will lead to a cascade in this region of momentum space. Outside this range, other dynamics will occur, which our description is agnostic to. More careful consideration of the cut-offs and corresponding boundary conditions seems needed if one would want to describe for example one cascade in a system that exhibits a dual cascade~\cite{PineiroOrioli:2015cpb}. A more correct procedure should take the dynamics beyond the grid boundaries into account and result in a scaling solution on the restricted grid that is consistent with the behavior on the full momentum range. We will leave this to future works and continue with boundary conditions that assume $f_s=0$ beyond cut-offs. Note that this assumption works well in the UV, as the suppressed tail does not contribute a lot to the relevant quantities even for the power-law tail. The result for the scaling function will be used as input in the calculation of the QNMs.

\vspace{5 pt}

\noindent \textbf{\emph{Far from equilibrium QNMs spectrum.--}} To find the spectrum of QNMs we discretize Eq.~\eqref{eq: eigenvalue equation} and solve it as an eigenvalue problem. Eq.~\eqref{eq: eigenvalue equation} needs some manipulations before it can be solved numerically. The leading behavior for $\delta f_\Omega$ is $\propto1/\bar{p}$ in the infrared and $\propto1/\bar{p}^4$ in the UV. To aid the numerics we take care of this by factoring out $1/(\bar{p}+\bar{p}^4)$ from both $f_s$ and $\delta f_\Omega$ and multiply the equation by $(\bar{q}+\bar{q}^4)$ since then $i\Omega\delta h_\Omega$ appears as a term and we keep the eigenvalue structure. We again work with the compact variable $u=\bar{p}/(1+\bar{p})$. The linearised operator on the right-hand side of the remaining equation diverges for $u\rightarrow0$. By multiplying both sides with $u$ this divergence is lifted and we changed the nature of the problem to a generalised eigenvalue equation, $M \vec{v}=\gamma N\vec{v}$, with $M$ and $N$ matrices, $\vec{v}$ na eigenvector and $\gamma$ the corresponding eigenvalue. In our case $N$ is given by the diagonal matrix with the values of our grid variable $u$ on the entries of the diagonal. 

\noindent Solving the generalised eigenvalue equation on the full compact grid $u\in[0,1]$ leads to perturbations that diverge for $u\rightarrow1$ compared to $f_s$. The assumption of linearity $\delta f_\Omega\ll f_s$ thus does not hold over an infinite momentum range and one should not consider the mode calculation for $\bar{p}\rightarrow\infty$. We can always ensure $\delta f_\Omega\ll f_s$ to hold until some momentum value since the amplitude of the perturbations can be chosen arbitrarily, note the linearity of the eigenvalue equation \eqref{eq: eigenvalue equation}. Complementary to our discussion of boundaries, w.r.t the assumption of overoccupation in the previous section, it seems necessary to include a UV cut-off in these mode calculations. 

\begin{figure}[h]
    \centering
    \includegraphics[width=0.8\linewidth]{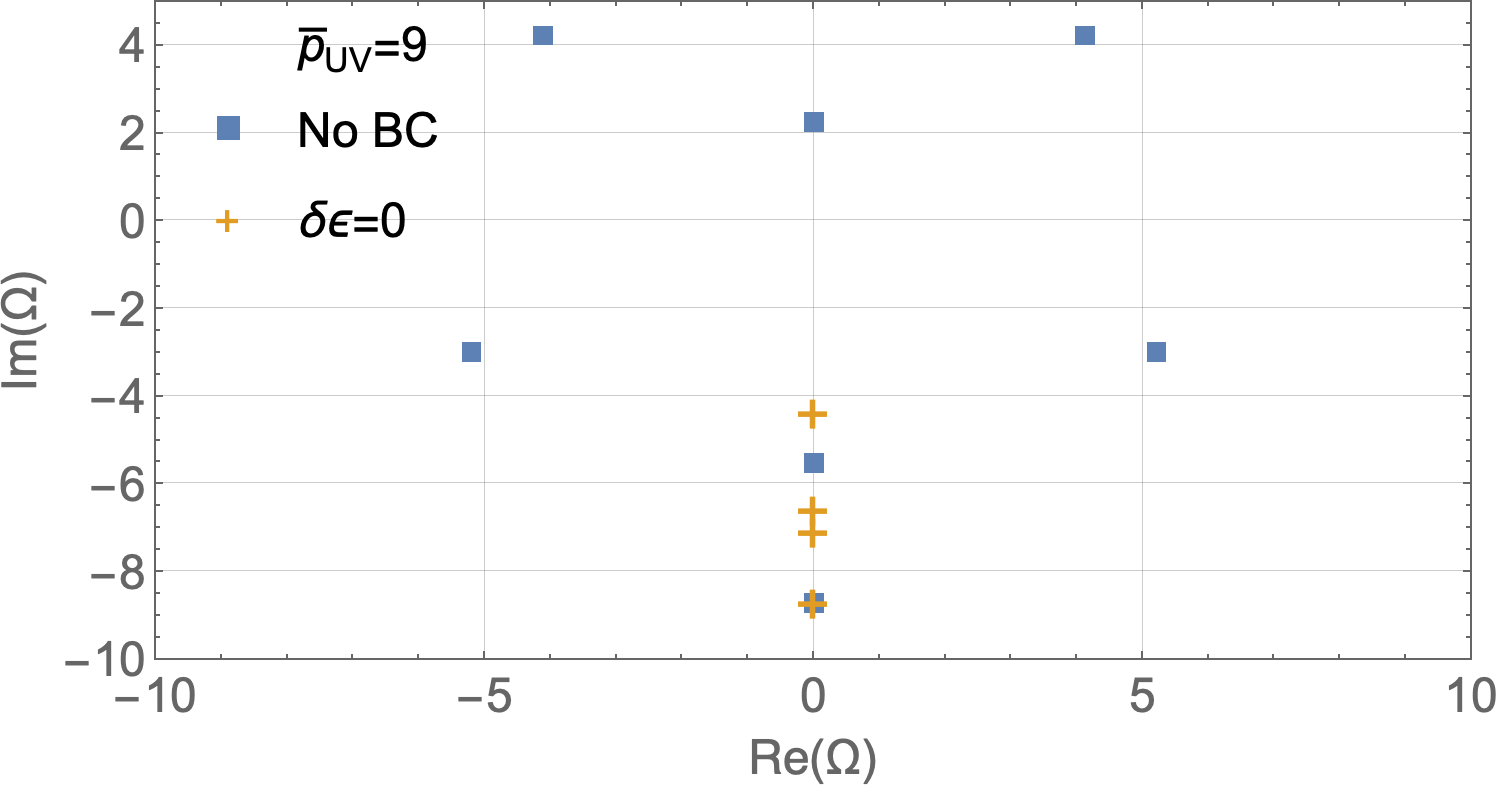}\\    \includegraphics[width=0.8\linewidth]{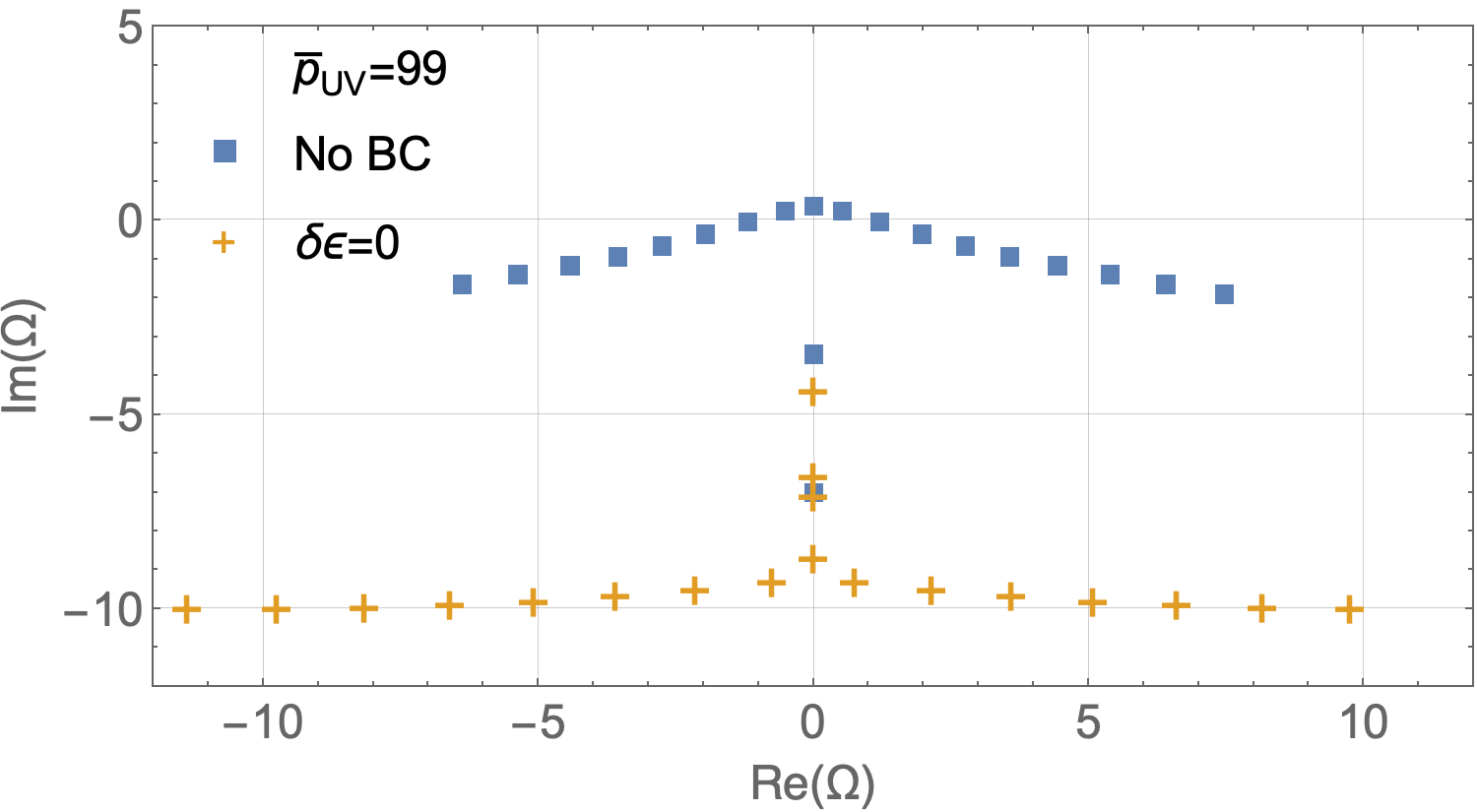}\\
    \includegraphics[width=0.8\linewidth]{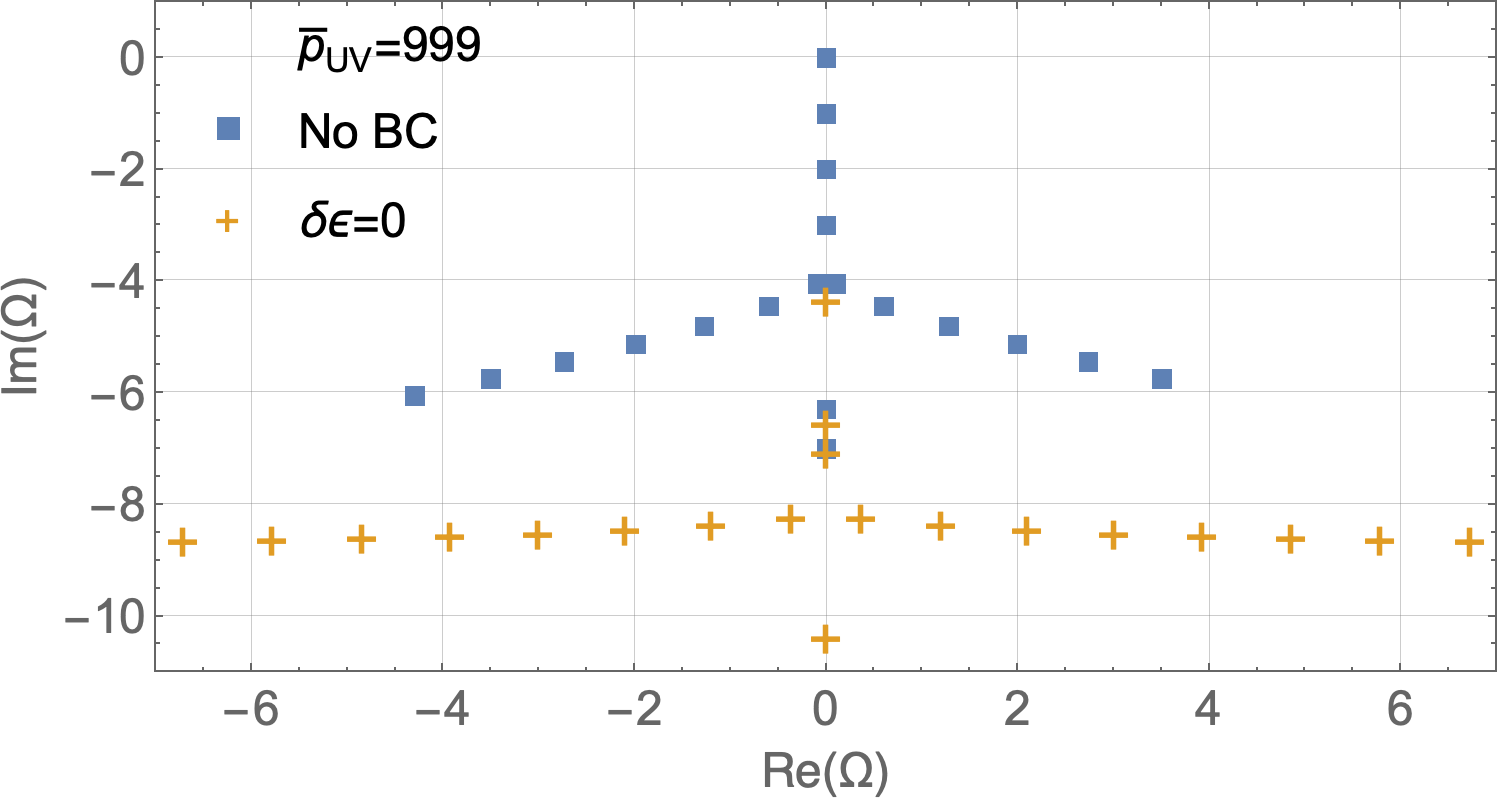}
    \caption{These plots show the results for the QNMs frequencies~$\Omega$ calculated on grids with different UV cut-offs $\bar{p}_{UV}$. When no boundary conditions are imposed and the cut-off is lowered there appear to be unstable modes, $\Im(\Omega)>0$. By imposing that the perturbation carry no energy density $\delta\epsilon
    =0$, all the modes are stable and become cut-off independent.}
    \label{fig: spectrum pUV 9}
\end{figure}

\begin{figure}[h]
    \centering
    \includegraphics[width=0.9\linewidth]{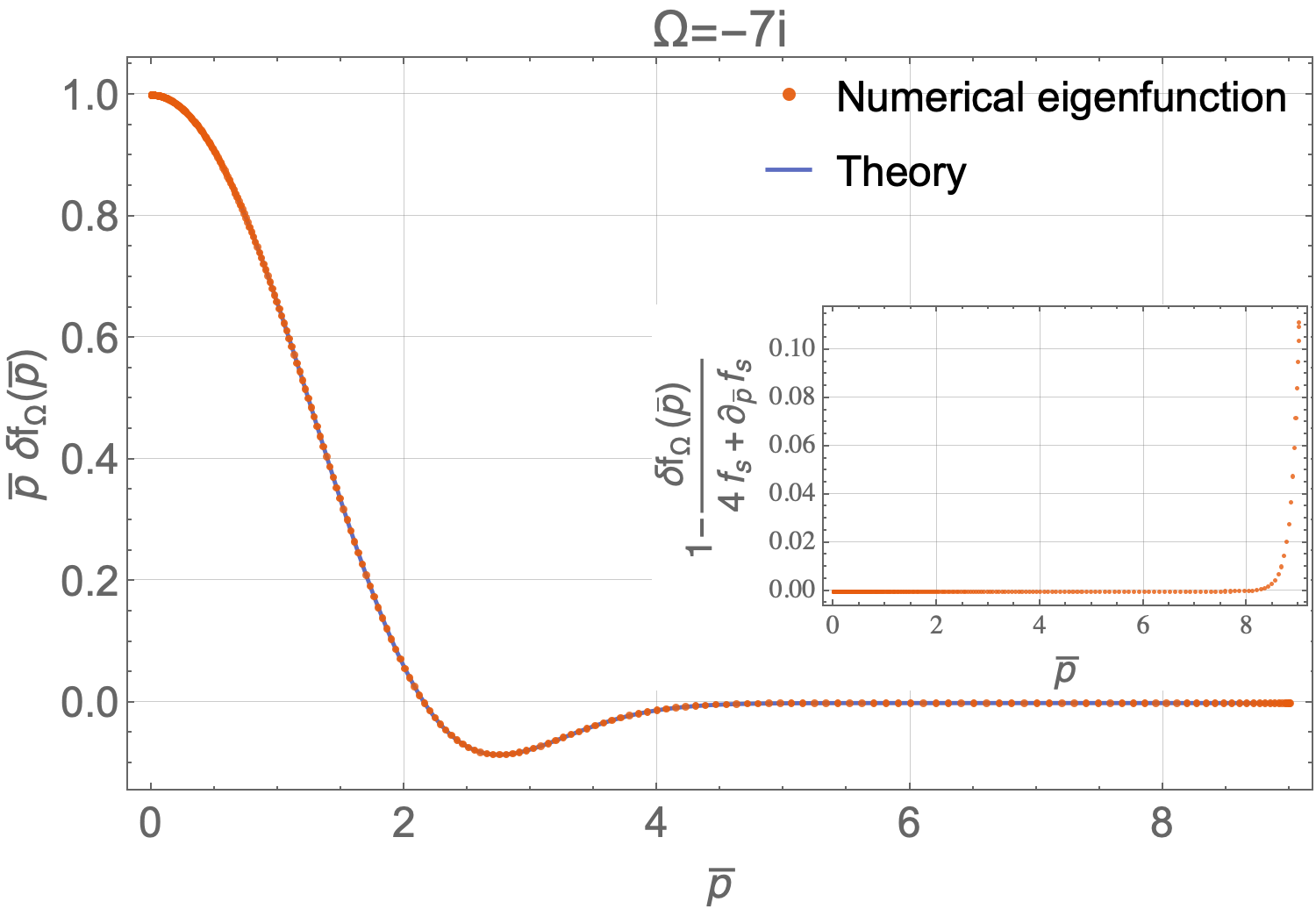}\\
    \includegraphics[width=0.9\linewidth]{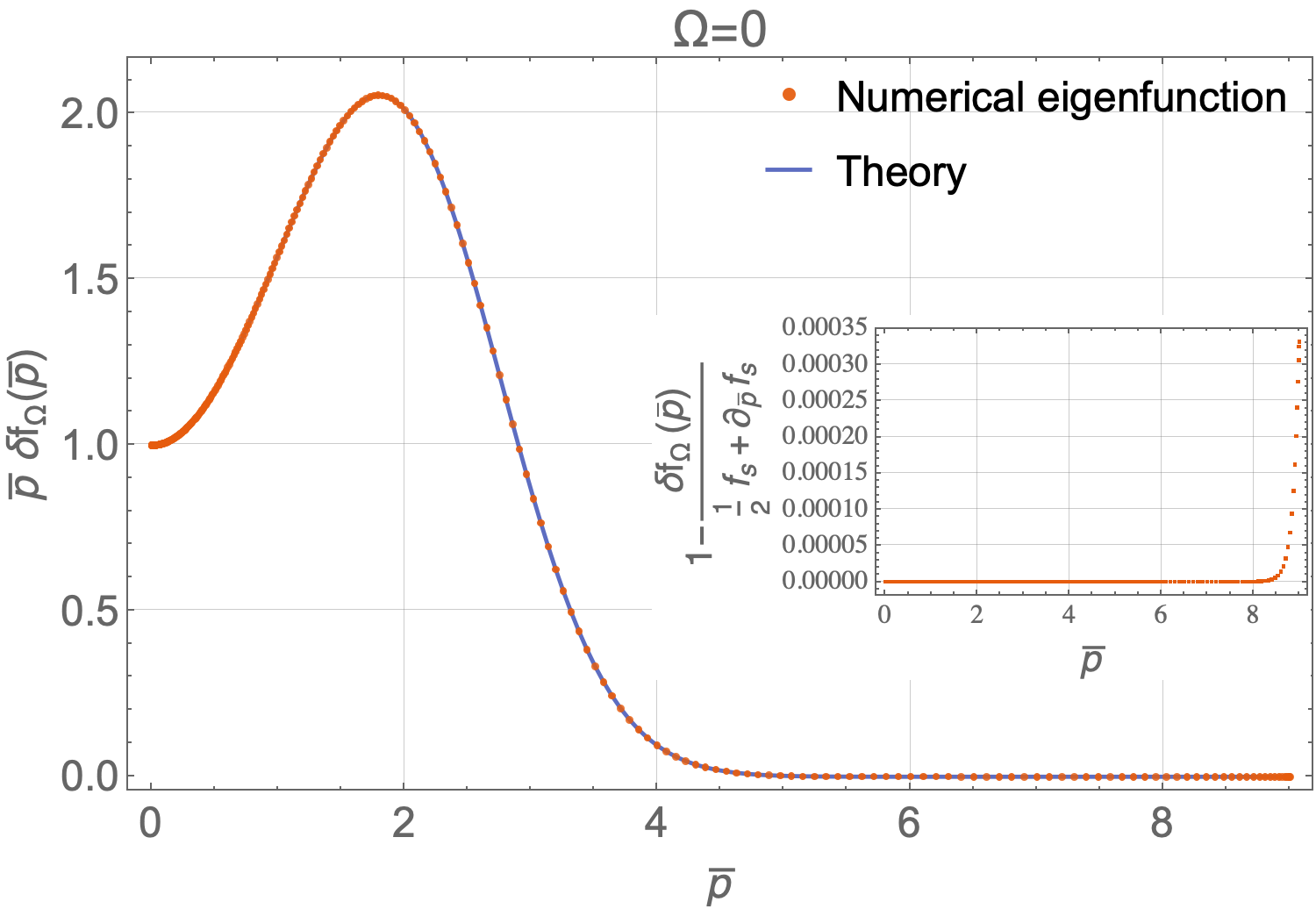}
    \caption{Comparisons of the momentum dependence of the modes from the eigenvalue calculation to their prediction in Eqs.~\eqref{eq: ts mode} and~\eqref{eq: tref mode}, respectively. In the inset their relative difference is shown. There is an excellent agreement up to the UV boundary where the numerical result starts to deviate from the prediction due to non-vanishing boundary effects.}
    \label{fig: ts and zero mode}
\end{figure}

Fig.~\ref{fig: spectrum pUV 9} shows the resulting spectra for different UV cut-offs $\bar{p}_{UV}$, while the IR cut-off is kept fixed at $\bar{p}_{IR}=0$. In Fig.~\ref{fig: spectrum pUV 9}(top and middle) there are seemingly unstable modes, $\Im{\Omega}>0$, when no boundary conditions are imposed, blue squares. Numerically, boundary conditions can be imposed by replacing one row in both matrices in the generalised eigenvalue equation with the linear discretized equation defining the relevant boundary condition. By imposing that the modes carry no energy density, $\delta\epsilon=0$, we get rid of the growing perturbations and the spectrum becomes that of an attractor. For the largest cut-off we consider, the structure is the same as for the other cut-offs. There seem to be only imaginary modes, note the predicted one at $\Omega=-7i$, see also Fig.~\ref{fig: ts and zero mode}(top). The other modes are at likely irrational imaginary numbers, $\Omega\approx-4.29i$, $\Omega\approx-6.49i$ and $\Omega\approx-8.59i$. They are very robust to changes in the UV cut-off and we checked that they remained stable when increasing the number of gridpoints. Finally, the horizontal discretized branch cut-like structure in Fig.~\ref{fig: spectrum pUV 9} (middle and bottom) moves down when increasing the number of gridpoints. We thus consider it to be a numerical artefact of the method.

\noindent The boundary condition also disposes of the predicted zero-mode $\Omega=0$. We calculate and compare the corresponding eigenfunction by setting $\Omega=0$ in the eigenvalue equation \eqref{eq: eigenvalue equation} and solving it as a linear equation for $\delta f_\Omega$. The result is shown in Fig.~\ref{fig: ts and zero mode}(bottom) and shows a very good agreement with the prediction, with the biggest deviation being near the UV cut-off due to boundary contributions. The theoretical form of the zero-mode $\frac{1}{2} f_s+\partial_{\bar{p}}f_s$ results from Eq.~\eqref{eq: tref mode} when taking into account that $t_{ref}$ appears as $1/\sqrt{t_{ref}}$ in the amplitude of the scaling function, see the discussion in the previous section.

\noindent We do not include spectra calculated with an IR cut-off since we discussed in the previous section that the $1/\bar{p}^4$ tail seems unphysical. A better understanding of how to treat the boundary conditions at this cut-off is needed to consider the eigenmode calculation. Mainly, we cannot impose the boundary condition $\delta\epsilon=0$, since the power-law tail sources large contributions to the boundary term in for example Eq.~\eqref{eq: energy conservation ts mode}. The assumption of conservation of energy density, with $\delta\epsilon=0$, only holds approximately. For the scaling functions with a Gaussian UV-tail, the values in the UV are suppressed and this approximation is justified. 

\vspace{5 pt}

\noindent \textbf{\emph{Outlook.--}} Our paper aims to provide a bridge between far from equilibrium NTFP and QNMs describing transient approach to thermality. It is natural to envision a significant advancement of our understanding about the attractive nature of NTFP and phenomena in their vicinity coming from adopting the relevant knowledge and techniques from the fields of black hole perturbation theory and holography.

There are many future directions that our work opens. On the theoretical front, it is certainly understanding the QNM spectrum for other collision kernels, in particular for the large-$N$ kinetic theory of encapsulating also the infrared cascade~\cite{Walz:2017ffj}. As a preliminary step in this direction, in Appendices~1,~2 and~4 we consider a toy model within the FK kinetic theory of infrared particle number conserving cascade and associated QNMs spectrum, which now contains also modes with nontrivial $\Re(\Omega)$. It would be also very interesting to connect our findings with earlier studies of stability of nonthermal fixed points using 2PI QFT techniques~\cite{Preis:2022uqs}. Finally, on the experimental front, it would be fascinating to directly observe QNMs of NTFP using their experimental realisations in cold atomic gases. A natural way to proceed would be to study deviations of the distributions function from the scaling form in the static frame, as encapsulated by Eq.~\eqref{eq: perturbation to prescaling bis}.

\begin{acknowledgments}
We would like to thank Alex Serantes and Robbe Brants for extensive discussions,  Jürgen Berges, Gabriel Denicol, Aleksas Mazeliauskas, Thimo Preis for discussions and collaborations on related topics and Cambridge University Quantum Gases and Collective Phenomena group, in particular Zoran Hadzibabic, Martin Gazo and Gevorg Martirosyan, for hospitality and in-depth discussions about experimental realizations of NTFPs. This project has received funding from the European Research Council (ERC) under the European Union’s Horizon 2020 research and innovation programme (grant number: 101089093 / project acronym: High-TheQ). Views and opinions expressed are however those of the authors only and do not necessarily reflect those of the European Union or the European Research Council. Neither the European Union nor the granting authority can be held responsible for them. 
\end{acknowledgments}

\bibliographystyle{bibstyl}
\bibliography{qnm_ntfp} 

\newpage

\appendix

\section{Appendices}

\noindent \textbf{\emph{Appendix 1: On the possibility of an infrared cascade.--}} The Fokker-Planck kinetic theory we consider describes elastic scatterings of gluons in the small-angle approximation. This means it should conserve both particle number density and energy density. In the main text we considered only energy conservation, which led to $\sigma=4$. If particle number is also conserved, this would result in $\sigma=3$, leading to a cascade characterized by the NTFP exponents $(\alpha,\beta)=(-3/5,-1/5)$.  FP kinetic theory is an approximation of QCD effective kinetic theory~\cite{Arnold:2002zm}. This kinetic theory breaks down for momentum values below the Debye mass $m_D\propto \lambda\int d^dp f/p$, with $\lambda$ the 't Hooft coupling constant and one can not trust the results of kinetic theory in this regime. If the IR also reaches a scaling form as in Eq.~\eqref{eq.NTFPdef}, the NTFP exponents would probably be different than the ones we find here since it has to be described by a different theory. Now we ignore this and investigate which dynamics are possible in FP kinetic theory. The Boltzmann equation with Eq.~\eqref{eq: CFP} conserves particle number density, if the particle flux $\mathcal{F}=-I_a p^2\partial_pf-I_bp^2f(f+1)$ vanishes at the boundaries in momentum space~\cite{Blaizot_2013}. At the UV boundary $p_{UV}$ this will be satisfied because physical distribution functions have finite support in momentum space. Specifically, $f$ should decay faster than $1/p^3$ to have a finite particle number density for $p_{UV}\rightarrow\infty$, in which case the flux will also vanish. Near the IR boundary $p_{IR}$ it is clear that if the leading behavior is constant $f\propto p^0$ the flux $\mathcal{F}$ vanishes for $p_{IR}\rightarrow0$. The other allowed behavior in the IR is $f\propto 1/p$, see \cite{Blaizot_2014} for a discussion on this. In this case the flux will only vanish if the coefficient of the leading $1/p$ behavior is $I_a/I_b$. Ab initio simulations often start from initial coniditions approximating a color-glass condensate $f(t=0,p)=\Theta(p/Q_s)$, with $\Theta$ the heavy-side function and $Q_s$ a saturation scale. The behavior in the IR is then initially constant. In the end these systems should thermalise and approach a Bose–Einstein distribution. This behaves as $\propto T/p$ in the IR. We see that the distribution function will build up a $1/p$ structure in the infrared. Now the flux $\mathcal{F}$ can only be zero at all times if $I_a/I_b$ starts at $0$ and approaches $T$ in thermal equillibium. The latter is true and can be seen by plugging in the Bose-Einstein distribution into the expressions for $I_a$ and $I_b$. On the other hand $I_a\neq0$ will always hold for $f(t,p)\geq0$ as should be the case for a well defined distribution function. We conclude that the flux can only remain zero if a discontinuous jump in the coefficient of the leading IR behavior $1/p$ occurs. This seems unphysical. In other words, the flux will be non-zero at times during the formation of $1/p$ behavior and will go to zero when thermal equillibrium is reached, meaning particle number can not be conserved throughout the full evolution. However, if the distribution is initialised at or reaches $p f(t,p)|_{p=0}=I_a/I_b$ during the evolution, it might be possible to keep particle number conserved for the remainder of the simulation, leading to a cascade. We check if the NTFP form Eq.~\eqref{eq.NTFPdef} is consistent with this. Let us assume the distribution function behaves as

\begin{equation}
    f=\frac{I_a}{I_b}\frac{1}{p}.
\end{equation}

\noindent  at leading order in the IR. The time dependence of $I_a$ and $I_b$ can be found by plugging Eq.~\eqref{eq.NTFPdef} into Eq.~\eqref{eq: integrals FP} and dropping the linear term in $f$ in $I_a$. The latter is justified if the overoccupied region where $f\gg1$ contributes significantly more than the region where $f\lessapprox1$. We find $I_a=A(t)^2B(t)^{-3}\tilde{I}_a$ and $I_b=A(t)B(t)^{-2}\tilde{I}_b$. Considering this in the equation above leads to 

\begin{equation}
    f=A(t)\frac{\tilde{I}_a}{\tilde{I}_b\bar{p}},
\end{equation}

\noindent where we used $p=B(t)^{-1}\bar{p}$. This has the same form as in Eq.~\eqref{eq.NTFPdef}, with $f_s=\frac{\tilde{I}_a}{\tilde{I}_b\bar{p}}$. At first sight it might thus be possible to realise a cascade characterized by particle number conservation. For the UV cascade everything is fine since it is not the flux itself, but $p \mathcal{F}$ that has to vanish at the boundaries, see \cite{Blaizot_2013}. This always holds for $p_{IR}\rightarrow0$, even when $f\propto1/p$ for small momenta.

\vspace{5 pt}

\noindent \textbf{\emph{Appendix 2: Analytical insights from scaling function for $\sigma=3$.--}} As discussed above there is the possibility that an IR cascade is realised in FP kinetic theory. We study what the scaling function $f_s$ would look like as a solution of the right-hand side equaility in Eq.~\eqref{eq: seperation of variables bis} with $\sigma=3$, $\beta=-1/5$ and the collision kernel from Eq.~\eqref{eq: CFP overoccupied}. Again we first rescale the momentum variable $\bar{p}$ and scaling function $f_s$ according to Eq.~\eqref{eq: subsitutions}. The resulting equation is 

\begin{equation}
    3 g_s(\bar{q})+\bar{q}\partial_{\bar{q}}g_s(\bar{q})=-\frac{1}{\bar{q}^2}\partial_{\bar{q}}\left(\bar{q}^2\partial_{\bar{q}}g_s(\bar{q})+\bar{q}^2g_s^2(\bar{q})\right)
    \label{eq: reduced eq gs}
\end{equation}

\noindent Interestingly, this equation has a closed form solution and we can use it to build intuition for the scaling function of the direct energy cascade with $\sigma=4$. The solution is complicated and can be found in the Mathematica notebook in the arXiv submission. It contains two integration constants, consistent with the second order nature of Eq.~\eqref{eq: reduced eq gs}. One integration constant has to be set to zero in order to have $\propto1/\bar{q}$ for $\bar{q}<<1$. This behavior was observed for the scaling function in \cite{AbraaoYork:2014hbk} and we start by looking for such solutions. The solution then has the following limiting behavior

\begin{equation}
\begin{split}
    &\text{$\bar{q}\gg1$: } c/\bar{q}^3+O(1/\bar{q}^5)\\
    &\text{$\bar{q}\ll1$: } d/\bar{q}+O(\bar{q})\\
\end{split}
\end{equation}

\noindent with $c$ the remaining integration constant and $d=(1+\sqrt{1+4c})/2$. For $c=0$, the leading behavior in the UV is $\sqrt{2/\pi}e^{-\bar{q}^2/2}$ and a Gaussian tail appears. In the IR the solution is then also completely fixed with $d=1$. This discussion shows that if we assume $1/\bar{q}$ near $\bar{q}=0$ and specify the value of $\bar{q}g_s(\bar{q})|_{\bar{q}=0}=const$, we have a unique solution. 

\noindent To come back on the discussion in Appendix 1, we see that the solution which has both integration constants set to zero is consistent particle number conservation. In that case, the coefficient of $g_s$ in the IR is $1$ and Eq.~\eqref{eq: subsitutions} shows that, by choosing $\tilde{I_a}=1/D_2$, the scaling function is indeed $\frac{\tilde{I}_a}{\tilde{I}_b}\frac{1}{\bar{p}}$ in the limit of $\bar{p}\rightarrow0$. The remaining question is if this scaling function is conistent with  the right-hand side equality in Eq.~\eqref{eq: seperation of variables bis}. 

\vspace{5 pt}

\noindent \textbf{\emph{Appendix 3: Numerical method to determine scaling function.--}} We calclate the scaling function using the Chebyshev-Gayuss-Lobatto pseudospectral method \cite{Boyd:Chebyshev}. This leads to a discretization of \eqref{eq: reduced eq gs}. As explained in the main text we solve the equation on a compact grid $u=\frac{\bar{q}}{1+\bar{q}}$. We first consider $u\in[0,1]$, so there are no boundaries in momentum space $\bar{q}$. To help the numerics we factor out ${1}/({\bar{q}+\bar{q}^3})$, which takes care of the leading behavior in the IR and UV. Doing this and substituting $\bar{q}\rightarrow \frac{u}{1-u}$ in Eq.~\ref{eq: reduced eq gs} leads to 

\begin{equation}
\begin{split}
    &\frac{1}{(1-2u+2u^2)^2}(4(-1+u)^3u h_s(u)^2\\&+2(-1+u)h_s(u)(-4u^3+(-1+u)^3(1-2u+2u^2)h_s'(u))\\&-u(1-2u+2u^2)((2-5u+8u^2-10u^3+4u^4)h_s'(u)\\&+(-1+u)^3(1-2u+2u^2)h_s''(u)))=0,\\
\end{split}
\end{equation}
 with $h_s(\frac{\bar{q}}{1-\bar{q}})=({\bar{q}+\bar{q}^3})g_s(\bar{q})$ and where we mulitiplied the equation by $\bar{q}(\bar{q}+\bar{q}^3)$ to get rid off overall prefactors that diverges for $u=0$ and $u=1$. To find $g_s$ we determine $h_s$ by an iterative procedure, which is sketched out as follows. Denote the equation above as $X[h_s]=0$. Now we can write $h_s=h_s^0+\delta h_s$ and expand $X[h_s^0+\delta h_s]$ to first order in $\delta h_s$. The resulting equation 

\begin{equation}
    X[h_s^0]+\frac{\delta X[h_s^0]}{\delta h_s^0}\delta h_s=0
    \label{eq: linearized equation for gs}
\end{equation}

\noindent is a linear equation for the unknown $\delta h_s$. Here, $\frac{\delta X[h_s^0]}{\delta h_s^0}$ has to be understood as an operator acting on $\delta h_s$ and is numerically represented by a matrix. Our procedure finds a solution by starting from an initial guess for $h_s^0$ and solving the matrix equation \eqref{eq: linearized equation for gs} for $\delta h_s$. Subsequently, the resulting $\delta h_s$ is added to $h_s^0$ and this procedure is repeated until the norm of $X[h_s^0+\delta h_s]$ is small enough. 

\noindent From the discussion in Appendix 2, it is clear that a solution is completely specified if we choose a value for $h_s$ at $u=0$ or alternatively at $u=1$ and keep this fixed throughout the iterative process. We impose this boundary condition by replacing a row in the matrix equation \eqref{eq: linearized equation for gs} with the correct numerical equation reflecting the boundary condition for $\delta h_s$, e.g. $(0,0,0,...,1) \delta h_s=c-h_s^0(u=1)$ to impose that $h_s(u=1)=c$ in the end. 

\noindent A self-consistent scaling function solution can be constructed from a solution $g_s(\bar{q})=\frac{1}{\bar{q}+\bar{q}^3}h_s(\frac{\bar{q}}{1-\bar{q}})$ for which $\tilde{I}_a[g_s]=\tilde{I}_b[g_s]$, see the section in the main text on the determination of the scaling function. We find such a solution by using a root finding procedure that searches for the coefficient $c=h_s(u=1)$ for which the condition is satisfied.

\begin{figure}
    \centering
    \includegraphics[width=0.9\linewidth]{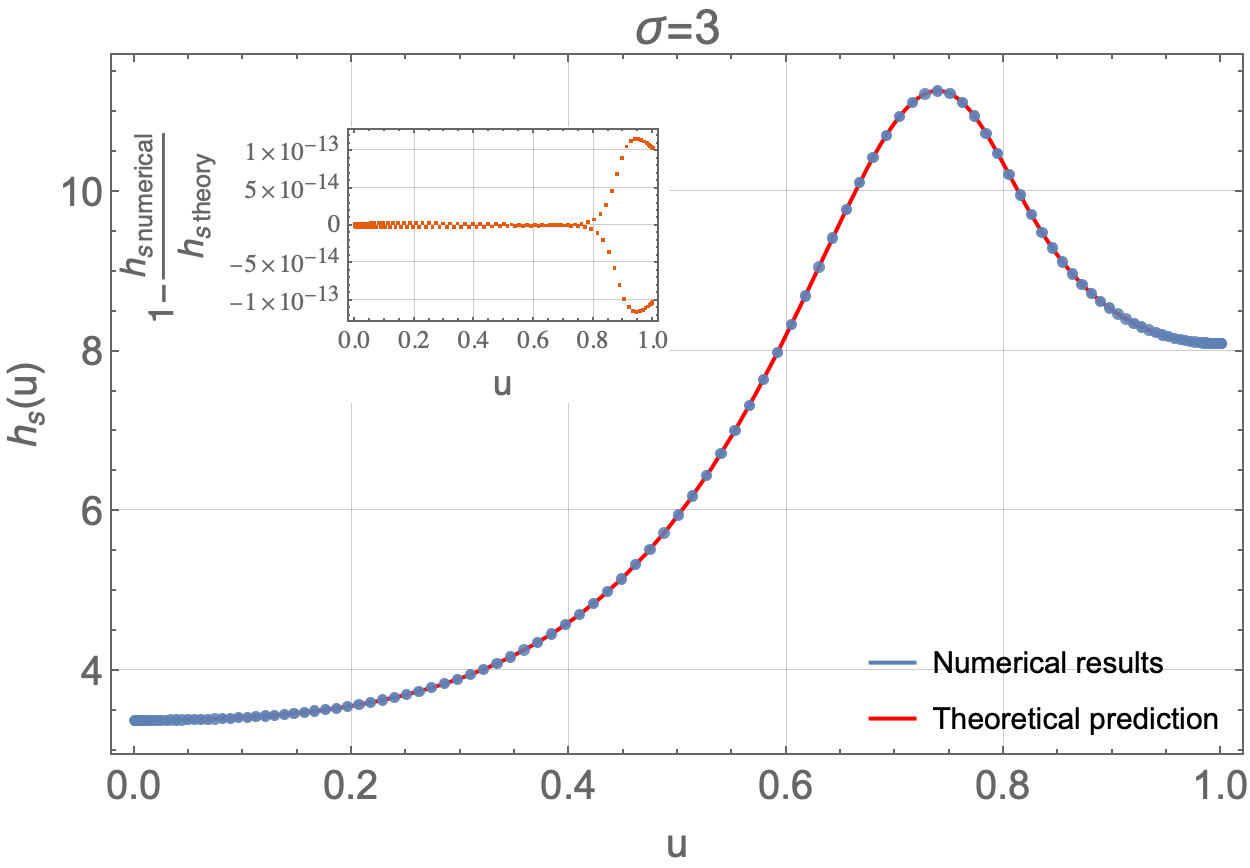}
    \caption{Result for $h_s(u)=({\bar{q}+\bar{q}^3})g_s(\bar{q})$ with $u=\bar{q}/(1-\bar{q})$. The result shown was calculated for $u\in[0,1]$. There is a very good agreement to the analytical solution, shown in red, also see the inset. In the analytical solution we set one integration to 0 and the other to the value of $h_s(u=1)$ of our numerical result, consistent with discussion in appendix 2.}
    \label{fig: scaling function sigma 3}
\end{figure}

\noindent The result for $h_s(u)$ for the case of $\sigma=3$ is shown in Fig.~\ref{fig: scaling function sigma 3}. As ansatz we took a constant, $h_s^0(u)=2$. The value for $u=1$ is $h_s(u=1)=8.106...$ and for $h_s(u=0)=3.39...$. If we reinstate the factored out behavior, this means the solution is $\propto1/\bar{q}$ in the IR and $1/\propto\bar{q}^3$ in the UV. The solution which would be consistent with particle conservation, the $h_s(u=0)=1$, does not satisfy $\tilde{I}_a[g_s]=\tilde{I}_b[g_s]$. This means that if there was only the cascade for particle conservation and $\sigma=3$, there would be no self-similar solution Eq.~\eqref{eq.NTFPdef}, since there is no consistent scaling function $f_s$. Since energy is also conserved, there will also be the UV cascade in the momentum region where the enregy density is dominant over the particle number density. It might then be possible, with the correct boundary conditions where both cascades are connected to each other, to find a consistent scaling solution. The investigation of this idea is left for future research.

\noindent The insights and ideas we descibed for the case $\sigma=3$ are transported to the case for $\sigma=4$, where no closed form expression for $g_s$ has been found. We consider Eq.~\eqref{eq: reduced eq gs} in which we replace 3 by 4. 

\begin{equation}
    4 g_s(\bar{q})+\bar{q}\partial_{\bar{q}}g_s(\bar{q})=-\frac{1}{\bar{q}^2}\partial_{\bar{q}}\left(\bar{q}^2\partial_{\bar{q}}g_s(\bar{q})+\bar{q}^2g_s^2(\bar{q})\right)
    \label{eq: reduced eq gs sigma 4}
\end{equation}

\begin{figure}[h]
    \centering
    \includegraphics[width=0.9\linewidth]{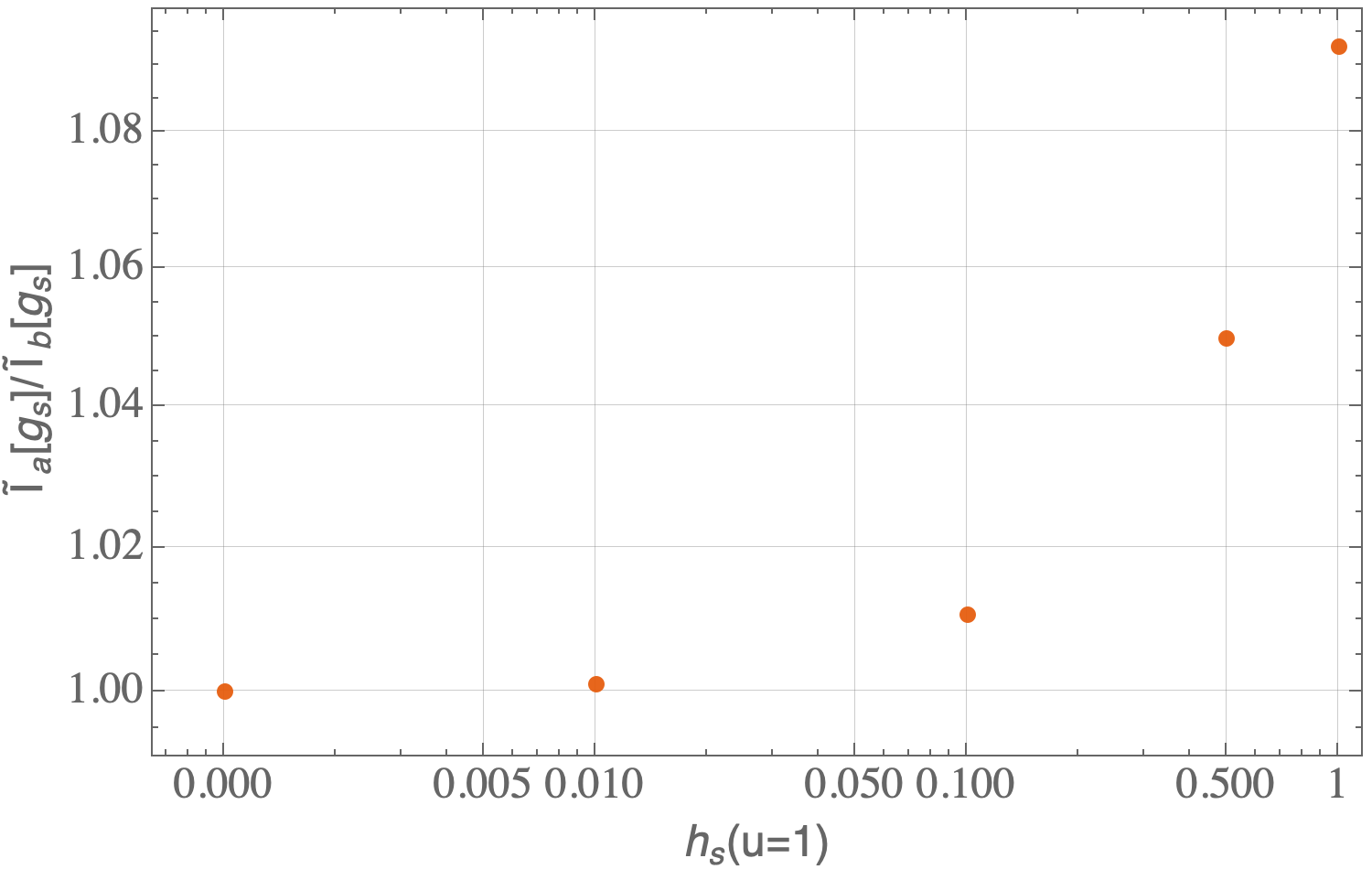}
    \caption{Different $g_s$ were calculated by fixing $h_s(u)=(\bar{q}+\bar{q}^4)g_s(\bar{q})$ at $u=\frac{\bar{q}}{1+\bar{q}}=1$ as a boundary condition. The ratio $\tilde{I}_a[g_s]/\tilde{I}_b[g_s]$ approaches 1 as the value of $h_s(u=1)$ is lowered to zero. This indicates the power-law tail gets killed and replaced by the remaining Gaussian tail to get a consistent solution.}
    \label{fig: UV coefficient sigma 4}
\end{figure}

\noindent The leading behavior in the IR is again $\propto1/\bar{q}$, but now $\propto1/\bar{q}$ in the UV. We thus factor out $1/(\bar{q}+\bar{q}^4)$ and repeat the procedure described above. Fig.~\ref{fig: UV coefficient sigma 4} shows how the ratio $\tilde{I}_a[g_s]/\tilde{I}_b[g_s]$ changes, when the value of $h_s(u=1)$ is changed. Clearly $\tilde{I}_a[g_s]/\tilde{I}_b[g_s]\rightarrow1$ in the limit of $h_s(u=1)\rightarrow0$. If the role of the integration constants for $\sigma=3$ is the same here, this means the power-law tail would be killed. Indeed, the tail of the solution on the $\bar{q}$-grid seems to go over into faster-than-power-law decaying behavior. To identify the non-power-law-like behavior at large $\bar{q}$ we set the nonlinear terms in \eqref{eq: reduced eq gs sigma 4} to zero since $g_s^2\ll g_s$ for a physical distribution function vanishing at infinite momentum. In this limit there is a closed form for the solution, which for $\bar{q}\gg 1$ behaves as $\frac{2\sqrt{2/\pi}c_1}{\bar{q}^4}+c_1O[1/\bar{q}^6]+e^{-\bar{q}^2/2}(\bar{q}(c_2-ic_1)+\frac{i c1 -c2}{\bar{q}})$, where $i$ is the well known imaginary number. We clearly see the power-law and Gaussian behavior. Setting $c_1$ to zero, leaves $e^{-\bar{q}^2/2}\bar{q}$ as dominant behavior for $\bar{q}\rightarrow\infty$. Fixing the value of $h_s$ at $u=1$ to be $0$ at this level, results in numerical issues since the gaussian makes the UV so small and small oscillations around zero appear close to $u=1$. To fix this we factor out $\frac{1}{\bar{q}}e^{-\bar{q}^2/2}(1+\bar{q}^2$. This takes care of both the IR and UV behavior. Now we repeat the iterative method described above to find the remaining momentum dependence. We can not arbitrarily choose the value in at $u=0$ or $u=1$ because all degrees of freedom have been taken care off by having $1/\bar{q}$ in the IR and killing the power-law $1/\bar{q}^4$ in the UV. There is only unique one solution which has the Gaussian tail and behaves as $1/\bar{q}$ for small momenta. The ratio $\tilde{I}_a[g_s]/\tilde{I}_b[g_s]$ becomes closer and closer to one when the number of gridpoints is increased. This implies that we have found the correct scaling function on the grid $u\in[0,1]$. The result, after reinstating the factored out behavior $\frac{1}{\bar{q}}e^{-\bar{q}^2/2}(1+\bar{q}^2$ and going back to the $\bar{q}=\bar{p}$ grid is shown in Fig.~\ref{fig: scaling function}.

\vspace{5 pt}

\noindent \textbf{\emph{Appendix 4: Technical discussion of eigenvalue problem.---}} We first consider the QNM calculation for the case of $\sigma=4$. To solve the eigenvalue equation Eq.~\eqref{eq: eigenvalue equation}, we use similar steps as in the determination of the scaling function. We discretize Eq.~\eqref{eq: eigenvalue equation} using the Chebyshev-Gayuss-Lobatto pseudospectral method \cite{Boyd:Chebyshev}. This makes the functional eigenvalue problem a standard finite matrix eigenvalue problem. We need to manipulate the equations a bit to be able to solve it numerically. We again factor out $1/(\bar{q}+\bar{q}^4)$, now from both $f_s=1/(\bar{q}+\bar{q}^4)h_s$ and $\delta f_\Omega=1/(\bar{q}+\bar{q}^4)\delta h_\Omega$, and multiply the equation by $(\bar{q}+\bar{q}^4)$ since then $i\Omega\delta h_\Omega$ appears as a term and we keep the eigenvalue structure. We go over to the compact variable $u=\bar{q}/(1+\bar{q})$. To get rid of numerically divergent terms for $u=0$ or $u=1$, we have to multiply the equation by $u$. Now we changed the nature of the problem to that of a generalised eigenvalue equation. This can be solved using known standard numerical methods.

\begin{figure}
    \centering
    \includegraphics[width=0.9\linewidth]{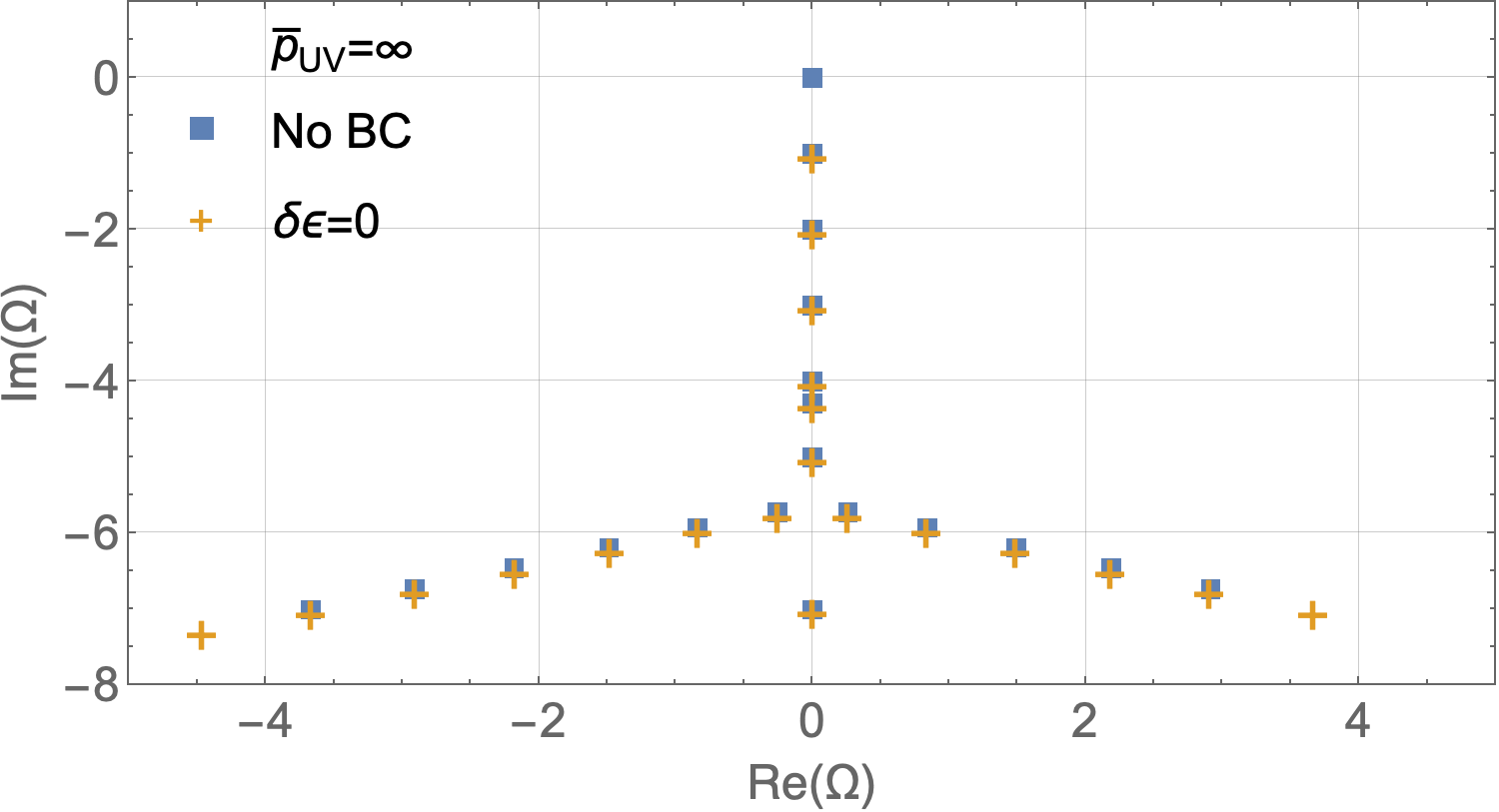}
    \caption{This figure shows the spectrum for the QNM calculation for $\sigma=4$. Imposing the boundary condition $\delta\epsilon=0$, does not change the spectrum except by getting rid of the zero-mode. We do not consider this spectrum to be physically realisable since $\delta f_\Omega\gg f_s$ in the UV. In the main text we took care of this by introducing a UV cut-off.}
    \label{fig: spectrum pUV inf}
\end{figure}

\noindent Here we show the spectrum calculated on the full grid in Fig.\ref{fig: spectrum pUV inf}. The structure here is interesting since it seems like there are the stable modes from before, see Fig.~\ref{fig: spectrum pUV 9}, with additional modes at every negative integer imaginary number. Imposing the boundary condition $\delta \epsilon=0$ does in this case not get rid of these modes. Compare this to the spectrum for UV cut-off $\bar{p}_{UV}=999$, see Fig.~\ref{fig: spectrum pUV 9} (bottom). It is not clear to us how this structure arises.

\begin{figure}
    \centering
    \includegraphics[width=0.9\linewidth]{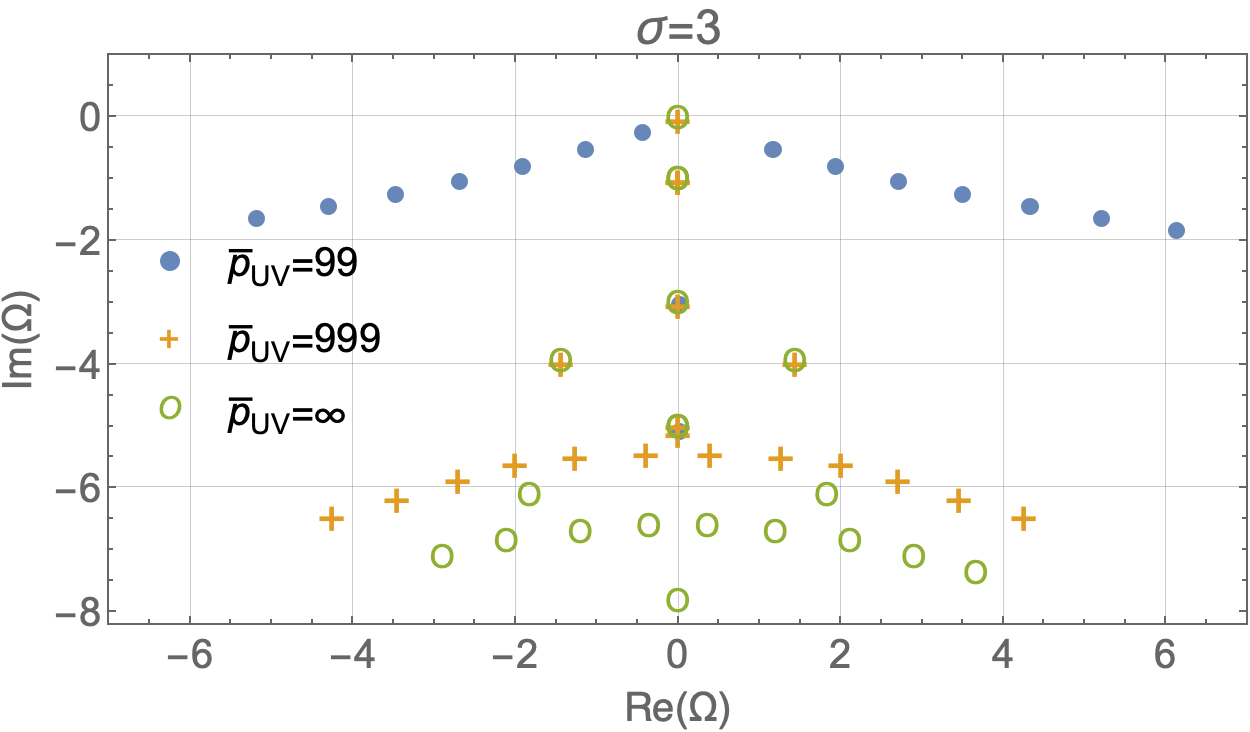}
    \caption{The spectrum of QNMs for a, maybe unrealisable, IR cascade in FP kinetic theory. The important feature are the modes around $-4i$ that have both an imaginary and real part.}
    \label{fig: spectrum sigma 3}
\end{figure}

\noindent Finally, we end this appendix by showing the result for a QNM calculation around the scaling function for $\sigma=3$, see Fig.~\ref{fig: spectrum sigma 3} without cut-offs. The method is the same as for $\sigma=4$ but with the different appropriate limiting behaviour factored out. As explained in Appendices~2 and~3, it seems that such a scaling function might not be realised without correctly taking the UV cascade into account. However, we are just interested in what the spectrum would look like if this was an allowed solution. Interestingly, there now seem to be stable modes that have both a real and imaginary part, very much reminiscent of QNM of black holes, dictating decaying oscillations. We do not consider the boundary condition imposed by the conservation law, $\delta n=0$ in this case from particle number conservation, since the boundary terms will have large contributions as explained below Eq.~\eqref{eq: energy conservation ts mode}.

\end{document}